\documentclass[pra,twocolumn,superscriptaddress,10pt,noshowpacs]{revtex4}
\usepackage[english]{babel}
\usepackage[T1]{fontenc}
\usepackage[utf8]{inputenc}
\usepackage{graphicx,epstopdf}
\usepackage{amssymb}
\usepackage{amsmath}

\usepackage{amsfonts}
\usepackage{bbm}
\usepackage{color}
\usepackage{latexsym}
\usepackage{caption}
\usepackage{subcaption}
\usepackage{times,txfonts}

\def\e{\mbox{e}}

\newcommand{\beq}{\begin{equation}}
\newcommand{\eeq}{\end{equation}}
\newcommand{\bea}{\begin{eqnarray}}
\newcommand{\eea}{\end{eqnarray}}
\begin{document}

\title{Generalized Ellis-Bronnikov graphene wormhole}

\author{T. F. de Souza\footnote{E-mail: felicio@fisica.ufc.br}}\affiliation{Universidade Federal do Cear\'a, Departamento de F\'{i}sica,  60455-760, Fortaleza, CE, Brazil.}

\author{A. C. A. Ramos\footnote{E-mail: antonio.ramos@ufca.edu.br}}\affiliation{Universidade Federal do Cariri, Centro de Ci\^encias e Tecnologia, 63048-080, Juazeiro do Norte, CE, Brazil.}

\author{R. N. Costa Filho\footnote{E-mail: raimundo.costafilho@gmail.com}}\affiliation{Universidade Federal do Cear\'a, Departamento de F\'{i}sica,  60455-760, Fortaleza, CE, Brazil.}

\author{J. Furtado\footnote{E-mail: job.furtado@ufca.edu.br}}\affiliation{Universidade Federal do Cariri, Centro de Ci\^encias e Tecnologia, 63048-080, Juazeiro do Norte, CE, Brazil.}

\date{\today}


\begin{abstract}

In this paper, we investigate the spinless stationary Schr\"odinger equation for the electron when it is permanently bound to a generalized Ellis-Bronnikov graphene wormhole-like surface. The curvature gives rise to a geometric potential affecting thus the electronic dynamics. The geometry of the wormhole's shape is controlled by the parameter $n$ which assumes even values. We discuss the role played by the parameter $n$ and the orbital angular momentum on bound states and probability density for the electron.

\end{abstract}

\maketitle


\section{Introduction}

The notion of a wormhole began by Flamm, Einstein, and Rosen \cite{flamm, einstein}, and was still developed by Wheeler \cite{wheeler}. It is known that wormholes emerges as a kind of solution of Einstein's field equations \cite{padmanabhan}. Through philosophical arguments, Weyl has speculated about these spacetime structures \cite{weyl, gibbons}, and for the original Einstein-Rosen solution, the wormhole throat does not allow the passage of classical objects. However, many argue that it could link quantum particles together to form entanglements \cite{pololsky, maldacena}.

From a topological point of view \cite{morris}, one can think of a wormhole as a tunnel connecting two asymptotically flat regions of the same universe or two different universes. One of the most important features of wormholes is the idea of traversability, first studied by Morris and Thorne \cite{morris}. Since the work of Morris and Thorne who stated that in order to construct a wormhole that is traversable one requires exotic matter as a source \cite{morris}. Therefore the pursuit for traversable wormholes in alternatives theories of gravity \cite{Sushkov:2005kj, Lobo:2005us, Garattini:2019ivd, Jusufi:2020rpw, Alencar:2021ejd, Oliveira:2021ypz, Carvalho:2021ajy, Richarte:2007zz, Matulich:2011ct, Richarte:2009zz, MontelongoGarcia:2011ag, Ovgun:2018xys, Lessa, Nilton:2022cho, Nilton:2022hrp} without the necessity of exotic matter is an intense topic of research. 

The first traversable solution for a wormhole was found by Ellis and Bronnikov \cite{ellis, bronnikov}. In his work, Bronnikov realized, with evidence, that the Ellis drainhole is geodesically complete, without event horizons, with free singularity and with traversability independent of direction \cite{ellis, bronnikov}. Besides, knowing that the wormhole’s scalar field source is phantom-like, then, all energy conditions of General Relativity (GR) are violated.

In the low energy physics, two-dimensional nanostructures, such as graphene \cite{geim, castro, katsnelson} and phosphorene \cite{carvalho}, have attracted attention due to their unusual properties. The electronic properties of such two-dimensional systems are highly dependent on the geometry \cite{dacosta, raimundo1, raimundo2}, so that they can be used as analogue models for high energy physics systems \cite{Capozziello:2020ncr, Alencar:2021ejd, Cvetic:2012vg, Pourhassan:2018wjg}. In addition, the effect of curvature in such two-dimensional systems opens the possibility of constructing new electronic devices based on curved graphene structures has motivated the study of graphene in several curved surfaces, such as M\"{o}bius-strip \cite{guo}, ripples \cite{juan}, corrugated surfaces \cite{saxena2010}, catenoid \cite{euclides, yesiltas, pdm-catenoid, helix}, Torus \cite{ramos, Yesiltas:2021crm, Yesiltas:2018zoy}, paraboloid \cite{paraboloid}, spheres \cite{dainam}, among others.

As a minimal surface, the two dimensional Ellis-Bronnikov wormhole geometry is equivalent to a catenoid \cite{euclides}. In Ref. \cite{gonzalez, pincak} a bridge connecting a bilayer graphene was proposed using a nanotube. In order to obtain a smooth connecting bridge, Ref.\cite{dandoloffsaxenajansen, dandoloff}, suggested a catenoid surface to describe the bilayer and the bridge using only one surface. This can be achieved due to the catenoid curvature which is concentrated around the bridge and vanishes asymptotically \cite{euclides}. For non-relativistic electrons, the surface curvature induces a geometric potential in the Schr\"{o}dinger equation. The effects of the geometry and external electric and magnetic fields upon the graphene catenoid bridge was explored in Ref.\cite{euclides}, where a single electron is governed by the Schr\"{o}dinger equation on the surface. Incidentally, the influence of a position-dependent mass problem upon the electron on a catenoid bridge was studied in Ref. \cite{pdm-catenoid}, where it was proposed an isotropic position-dependent mass as a function of the Gaussian and mean curvatures.

In this paper, we investigate the spinless stationary Schr\"odinger equation for the electron when it is permanently bound to a generalized Ellis-Bronnikov graphene wormhole-like surface. The generalized Ellis-Bronnikov wormhole is characterized by a function controlling discrete deformations from a catenoid to a cylinder. The curvature gives rise to a geometric potential affecting thus the electronic dynamics. We discuss the role played by the parameter $n$ that controls the deformation and the orbital angular momentum on bound states and probability density for the electron.

This article is divided as follows: in section II, we discuss elements of the generalized Ellis-Bronnikov wormhole in order to obtain the time-independent Schrodinger equation for the spinless electron; in section III, we study in details the electron behaviour from the geometric potential induced by the surface. The numerical results for the bound states are obtained and discussed in section IV; in section V the final remarks are outlined.


\section{Generalized Ellis-Bronnikov wormhole spacetime}

A generic Morris-Thorne wormhole is a Lorentzian manifold whose line element can be written as
\begin{equation}
    ds^2=-e^{2\Phi(r)}dt^2+\frac{dr^2}{1-b(r)/r}+r^2d\Omega_2, \label{Eq1}
\end{equation}
which is in the lorentzian signature ($-$ $+$ $+$ $+$), where $e^{2\Phi(r)}$ is the redshift function and $b(r)$ is the shape function. The $(r, \theta, \phi)$ are spherically polar coordinates and $d\Omega_2=d\theta^2+\sin\theta d\phi^2$.

For the usual Ellis-Bronnikov wormhole we have $\Phi(r)=0$ and $b(r)=R^2/r$, charectarizing a zero tidal wormhole with $R$ being the throat radius. However, a generalized version of the the usual Ellis-Bronnikov wormhole was proposed in \cite{kar, roy} satistying all the necessary Morris-Thorne conditions for making Lorentzian traversable wormholes. This generalized Ellis-Bronnikov (GEB) wormhole is characterized by two parameters, namely $n$ and $R$, that tell us about its size and shape. Then the GEB wormhole line element is given by
\begin{equation}
    ds^2=-dt^2+du^2+f^2(u)d\Omega_2, \label{Eq2}
\end{equation}
with $f(u)=\left(u^n+R^n\right)^{1/n}$ being a well-behaved function for $n$ even integers. Hence the GEB line element can be written as \cite{roy, morris}
\begin{equation}
	ds^2=du^2+\left(R^n+u^n\right)^{2/n}d\Omega_2, \label{1}
\end{equation}
where $u=\left(r^2-R^2\right)^{1/2}$ is proper radial distance coordinate (tortoise). The cylindrical angular coordinate $\phi \in[0,2\pi)$ is called parallel. These coordinates are part of a parallel-meridian cartesian system that is capable of covering the entire space of the GEB wormhole. In these coordinates $-\infty < u <\infty$, which is different from the cylindrical radial coordinate $\rho$ because $0 \leq \rho < \infty$, allowing us to distinguish between the lower and upper layers of the graphene, and $n$ is even integer. By considering an slice $\theta = \pi/2$, the line element can be rewriten as
\begin{equation}
    ds^2=du^2+\left(R^n+u^n\right)^{2/n}d\phi^2. \label{3}
\end{equation}
From (\ref{3}), the non-null components of metric tensor $g_{ij}$ ($i,j=1,2$) are given by
\begin{align}
	    g_{uu}(u)&=1, \label{4} \\
    	g_{\phi \phi}(u)&=\left(R^n+u^n\right)^{2/n}. \label{5}
\end{align}
The non-vanishing components of the Christoffel symbols \cite{padmanabhan, wald, manfredo, spivak} $\Gamma^j_{ik}=\frac{g^{jm}}{2}(\partial_ig_{mk}+\partial_kg_{mi}-\partial_mg_{ik})$ are straightforwardly calculated and written as
\begin{align}
    \Gamma_{\phi \phi}^u(u)&=-u^{n-1}\left(R^n+u^n\right)^{2/n-1}, \label{6} \\
    \Gamma_{u\phi}^\phi(u)&=\Gamma_{\phi u}^\phi(u)=\frac{u^{n-1}}{R^n+u^n}. \label{7}
\end{align}

Let us now consider an electron of effective mass $m^*$ and electric charge $-e$ permanently constrained to the surface of a GEB wormhole surface due to a geometric potential $V_\textrm{g}$, as given by the following Hamiltonian, 

\begin{align}
    \hat{H}&=\frac{1}{2m^*}g^{ij}\hat{P}_i\hat{P}_j+V_\textrm{g}.
\end{align}
The momentum operator is writen as $\hat{P}_i=-{\rm i}\hbar \nabla_i$, where the electron couples with the surface by means of the induced metric of the GEB wormhole surface $g_{ij}$ and the covariant derivative $\nabla_iV^j=\partial_iV^j+\Gamma^{j}_{ik}V^k$.  

In addition we consider also the da Costa's potential \cite{dacosta} $V_\textrm{dC}=-\frac{\hbar^2}{2m^*}\left(H^2-K\right)$, where $H$ is the mean curvature and $K$ is the Gaussian curvature. The da Costa's potential yields for the GEB wormhole surface the following expression (See appendix)
\begin{align}
    V_\textrm{dC}(u)&=-\frac{\hbar^2}{2m^*}\frac{(n-1)R^nu^{n-2}}{(u^n+R^n)^{2}} \nonumber \\
    &\hspace{4mm}-\frac{\hbar^2}{8m^*}\left(\frac{\left[1-u^{n-2}(u^n+R^n)^{2/n-2}\left(u^n+(n-1)R^n\right)\right]^2}{(u^n+R^n)^{2/n}[1-u^{2n-2}(u^n+R^n)^{2/n-2}]}\right). \label{33}
\end{align}
For $n=2$ we recover the da Costa potential for the catenoid previously addressed in the literature \cite{dandoloffsaxenajansen, euclides}.

The wave function, because of the axial symmetry (about the $z$-axis) of the GEB wormhole surface, must be invariant under the transformation of U(1), that is,
\begin{equation}
	\Psi(u, \phi)=\Phi(u)\,\e^{{\rm i}m\phi}, \, \, m=0, \pm 1, \pm 2, ...\, . \vspace{1mm}
\end{equation}

Therefore, the spinless and stationary Schr\"odinger equation for the electron, can be expressed as
\begin{align}
    -\frac{\hbar^{2}}{2m^{*}}\Phi''(u)&-\frac{\hbar^{2}}{2m^{*}}\frac{u^{n-1}}{u^n+R^n}\, \Phi'(u) \nonumber \\
    &+\left(\frac{\hbar^{2}}{2m^{*}}\frac{m^2}{\left(u^n+R^n\right)^{2/n}}+V_\textrm{dC}(u)\right)\Phi(u)=\varepsilon\, \Phi(u), \label{36}
 \end{align}
where $\varepsilon$ are the eigenvalues of $\hat{H}\Psi(u,\phi)=\varepsilon \Psi(u,\phi)$. In what follows we will discuss some implications of the geometry of the GEB wormhole surface on the electron dynamics.

\begin{figure*}
    \centering
    \includegraphics[scale=0.9]{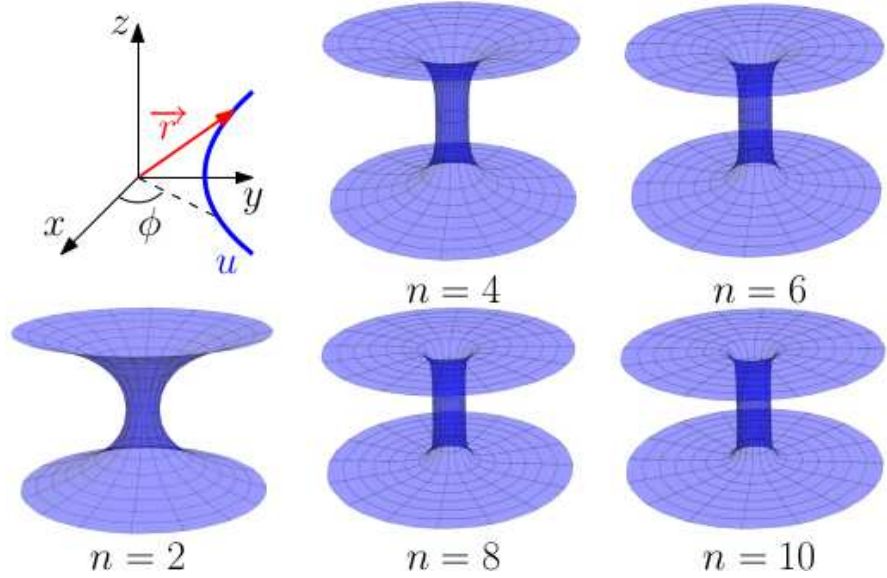}
    \caption{Above, the coordinate system, the vector, $\vec{r}=f(u)\cos{\phi} \ \hat{i}+f(u)\sin{\phi} \ \hat{j}+h(u) \ \hat{k}$, locates any point on the meridian $u$ with respect to the origin of the coordinate system. For $n=2$, the conventional Ellis-Bronnikov wormhole surface, which has the catenoid geometry. For $n=4, 6, 8$ and $10$, several the generalized Ellis-Bronnikov wormhole surfaces are presented. These surfaces tend to cylindrical geometries as $n$ increases. Here the radius, $R$, is 20 \AA.}
    \label{superficie}
\end{figure*}


\section{Geometry effects}

The equation (\ref{36}) governs the dynamics of the electron on the GEB wormhole surface. At this point it is important to highlight here that the first order derivative term is not hermitian, since
\begin{align}
         \left\langle \Phi_1 \biggl| -\frac{u^{n-1}}{u^n+R^n} \frac{\hat{P}_u}{{\rm i} \hbar}\, \Phi_2 \right\rangle &=-\left\langle -\frac{u^{n-1}}{u^n+R^n}\frac{\hat{P}_u}{{\rm i}\hbar}\, \Phi_1 \biggl| \Phi_2\right\rangle \nonumber \\
         &\hspace{2mm} +\left\langle \Phi_1\frac{u^{2n-2}-(n-1)R^nu^{n-2}}{(u^n+R^n)^2} \bigg| \Phi_2\right\rangle, \label{37}
\end{align}
where $\hat{P}_u=-i\hbar\partial_u$. However the Hamiltonian associated with equation (\ref{36}) is symmetric under simultaneous application of parity and time reversal operators, i.e., $\mathcal{PT}\hat{H}\mathcal{PT}=\hat{H}$. Thus, since the space-time reflection symmetry is preserved, the spectrum of the Hamiltonian eigenvalues is entirely real even though it is not hermitian \cite{lekner, bender, boettcher, jones, andrianov2007, joshi, andrianov1982}. 
Therefore, it is possible to find an Hermitean equivalent Hamiltonian possessing the same eigenvalue spectrum. In order to do so, let us perform the following change of variable
\begin{equation}
    \Phi(u)=\left(u^n+R^n\right)^{-1/(2n)}y(u), \label{38}
\end{equation}
with $y(u)$ satisfying a Schr\"{o}dinger-like equation, so that (\ref{36}) becomes
\begin{equation}
    -\frac{\hbar^2}{2m^*}y''(u)+V_\textrm{eff}(u)y(u)=\varepsilon y(u),\label{39}
\end{equation}
with
\begin{align}
    V_\textrm{eff}(u)=V_\textrm{ind}(u) +V_\textrm{dC}(u), \label{40}
\end{align}
being that
\begin{align}
V_\textrm{ind}(u)=-\frac{\hbar^2}{2m^*}\left(\frac{u^{n-2}\left((2-2n)R^n+u^n\right)}{4(u^n+R^n)^{2}} -\frac{m^{2}}{(u^n+R^n)^{2/n}}\right).
\label{41a}
\end{align}
As we can see, the effective potential is composed of the induced potential by the surface on the electron, $V_\textrm{ind}(u)$, which is essentially a repulsive potential, and by the da Costa potential, $V_\textrm{dC}(u)$, which emerges from the interaction of the electron with the graphene surface, being an attractive potential. These potentials, $V_\textrm{ind}(u)$ and $V_\textrm{dC}(u)$, compete with each other when the parameters of the Ellis-Bronnikov wormhole are changed, such as the radius $R$ of the wormhole, or when the wormhole geometry is changed. 

Therefore the behavior of the electron confined to the surface of the GEB wormhole made of graphene is described by an Hermitian Hamiltonian with an effective potential written as (\ref{40}).

In the limit $u \rightarrow \pm \infty$, we have free asymptotic states for the electron, because $y''+k^2y \cong 0$ ($k^2 = 2m^* \varepsilon/\hbar^2$, which is positive to allow scattered free asymptotic states), where it is clearly seen that the effective potential cancels out. And for $R \rightarrow 0$, the solution is proportional to the Bessel function of the first type \cite{abramowitz}. In summary, when (i) $u \rightarrow \pm\infty$, $y(u)=A\cos(ku+\varphi)$ ($A$ is an amplitude for $y$ and $\varphi$ is an initial phase) and (ii) $R \rightarrow 0$, $y(ku)=NJ_m(ku)$ ($N$ is a normalization constant for $y$). These asymptotic results are identical to those obtained for $n =2$. So far, in these regimes, the graphene nanophysics does not depend on $n$.

\begin{figure}[h]
    \centering
    \includegraphics[scale=0.6]{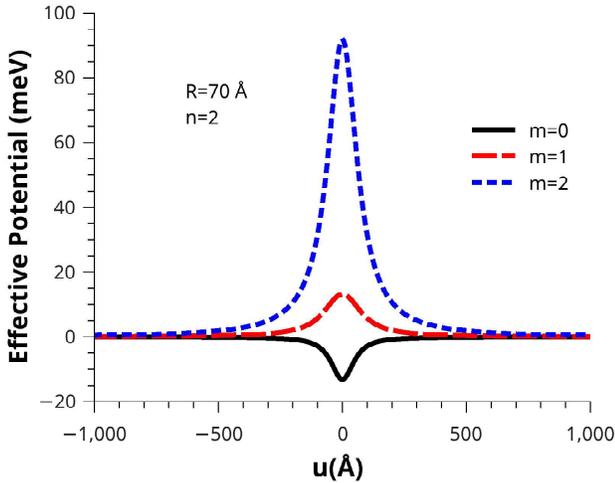}
    \caption{Effective potential for $R=70 $ \AA \ and $n = 2$, for some values of $m$.}
    \label{g1}
\end{figure}

Before discussing the bound states for an electron on the surface of the GEB graphene wormhole, we need to discuss the effective potential generated by this surface. For $n=2$, the GEB wormhole recovers the conventional Ellis-Bronnikov wormhole, with the wormhole having the geometry of a catenoid, as shown the Fig.\ref{superficie}. For this geometry the effective potential is shown in Fig.\ref{g1}. The solid black, dashed red and blue dotted lines correspond to orbital angular momentum, $m=0, 1$ and $2$, respectively, for $R=70 \, \, \AA$ and $m^*=0.03 m_0$ \cite{stabile}. This configuration was already addressed in the literature \cite{euclides}. 

For the orbital angular momentum $m=0$, the effect da Costa potential, which is attractive, is more pronounced than the surface-induced potential, which is repulsive, so the effective potential takes the form of a potential well centered, at the origin of the wormhole, in $ u=0$, at this point the value of the potential is $-13.12$ meV (see the Fig.\ref{g1}). However, when the orbital angular momentum is taken into account, i.e. when $m \neq 0$, the induced potential is drastically modified, due to the centrifugal term, in (\ref{41a}), and under these conditions, the effect of the induced potential becomes greater than the effect produced by the da Costa potential, so the effective potential takes the form of a potential barrier, also centered on the origin of the wormhole, according to Fig.\ref{g1}. These potentials are widely discussed in \cite{euclides}.

\begin{figure}[h]
    \centering
    \includegraphics[scale=0.6]{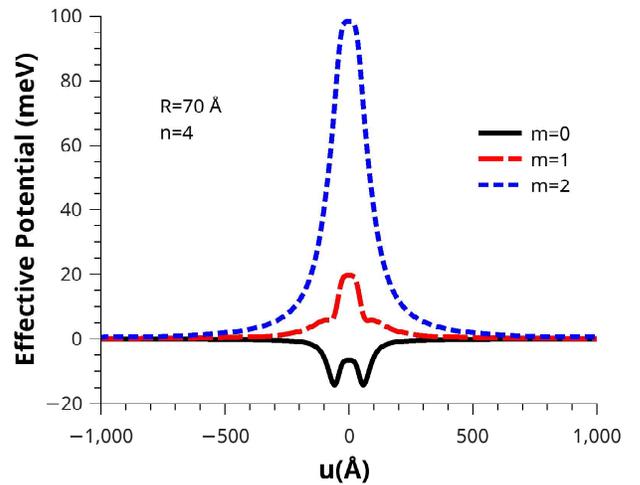}
    \caption{Effective potential for $R=70 $ \AA \ and $n = 4$, for some values of $m$.}
    \label{g2}
\end{figure}

In the Fig.\ref{g2}, the wormhole geometry is changed for $n=4$, as it changes the effective potential. For $m=0$, the da Costa's potential is more significant than the induced potential, but now the effective potential takes the form of a double well, whose minima are located at $u=-58\,\, \AA$ and $u=58 \, \, \AA$, and at these points, the value of the potential is $-14.2$ meV, as shown in Fig.\ref{g2}. This potential in the form of a double well arises because the generalized Ellis-Bronnikov wormhole geometry presents two points of more intense curvature, one at each end of the wormhole (see the figure \ref{superficie}). For $m\neq 0$, the induced potential becomes more relevant than the Costa's potential, due to the centrifugal term, and the effective potential takes the form of a potential barrier, located at the origin of the wormhole, as shown in the Fig.\ref{g2}.\\

\begin{figure}
    \centering
    \includegraphics[scale=0.6]{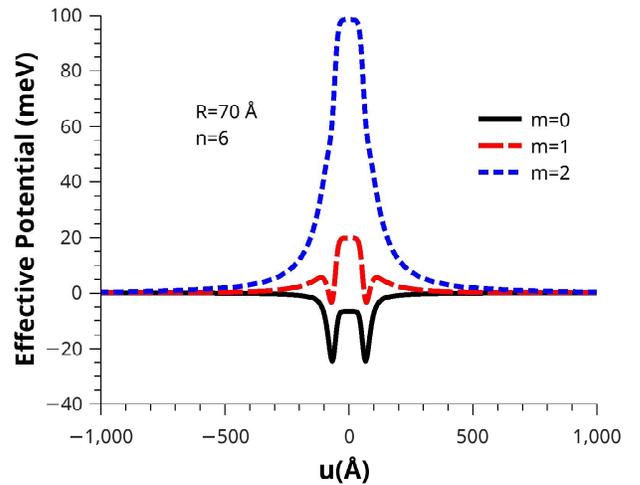}
    \caption{Effective potential for $R=70 $ \AA \ and $n = 6$, for some values of $m$.}
    \label{g3}
\end{figure}

The discussion of effective potential shown in Fig.\ref{g3} is qualitatively similar to the effective potential shown in Fig.\ref{g2}. However, the double potential well shown in Fig.\ref{g3} is deeper than the double potential well shown in Fig.\ref{g2}. As the surface of the generalized Ellis-Bronnikov wormhole assumes the shape of a cylinder, for larger values of $n$ (see figure \ref{superficie}), the effect of the curvature of graphene at the ends of the wormhole is accentuated, that explains why the depth of the double well is greater in Fig.\ref{g3}, when $n=6$, than in Fig.\ref{g2}, when $n=4$. The two minima of the effective potential, for $m=0$, shown in Fig.\ref{g3}, are located at $u=-67\, \, \AA$ and $u=67 \, \, \AA$, and at these points the value of the potential is $-24.4$ meV.

\section{Bound states}

In the previous section we studied the interaction of an electron on various types of generalized Ellis-Bronnikov wormhole surfaces, of radii $R$, made of graphene. The effective potential was obtained from this study given by (\ref{40}). In this section we will study their bound states, considering the radius of these surfaces equal to $R=70\,\, \AA$, and the effective mass of the electron in graphene equal to $m^{*}=0.03 m_{0}$.
For this, we solve numerically the (\ref{39}), using the finite
difference method \cite{ramos2}, for the effective potential given by the (\ref{40}).

\begin{figure*}
    \centering
    \includegraphics[scale=1.0]{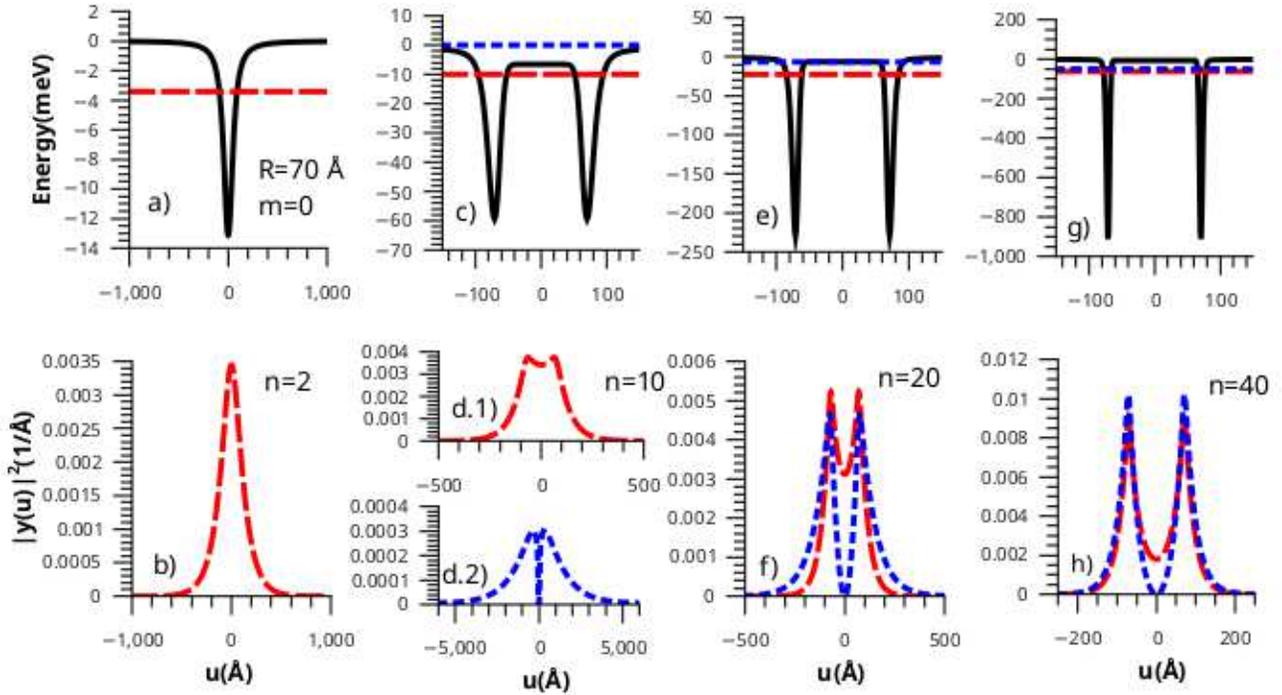}
    \caption{The bound states and their probability densities for a generalized Ellis-Bronnikov wormhole with radius $R=70$ \AA, \ $m=0$ and $m^{*}=0.03m_{0}$. The solid black line represents the effective potential for: a) $n=2$, c) $n=10$, e) $n=20$ and g) $n=40$. The dashed red and dotted blue lines correspond correspond to the ground state and first excited state and their probability densities, respectively.}
    \label{g4}
\end{figure*}

The Fig.\ref{g4} shows four effective potentials associated with four GEB wormhole surfaces made of graphene, namely a) $n=2$, c) $n=10$, e) $n=20$ and g) $n=40$. In these potentials the orbital angular momentum is not taken into account, so $m=0$. In Fig.\ref{g4} a), the effective potential for an electron confined to the surface of a conventional Ellis-Bronnikov wormhole is shown by the solid black line. For this potential there is a single confined state, whose energy is $-3.42$ meV. Its probability density function is a Gaussian function, whose the width of the half height is $\Delta u=231\, \, \AA$, as shown in Fig.\ref{g4} b). Making use of the angular symmetry of the system and taking the width of the half height of the probability density function is possible visualize a probability cloud, centered at the origin of the surface wormhole, in the form of a ring, or probability ring, where the electron is most likely to be found.     

For $n=10$, the effective potential, shown in Fig.\ref{g4} c), has the shape of a double potential well (solid black line), whose minima are located at $u=-70.5\, \, \AA$ and $u=70.5\,\, \AA$, with value of $-59.3$ meV. The dashed red and blue dotted lines represent the ground state and the first excited state, respectively, being that the energies of these states are $-9.98$ meV and $-0.034$ meV. These states are said to be hybrids, since the effective potential is composed of two potential wells that are close together and their states combine and form hybrid states \cite{novaes}. This can be seen in the probability density functions shown in figures \ref{g4} d.1) and \ref{g4} d.2). The Fig.\ref{g4} d.1) shows the probability density function of the ground state of the system, which has the form of two practically overlapping Gaussian functions, whose maxima are located at $u=-64\,\,\AA$ and $u=64\,\, \AA$. The width of the half height of this probability density function is $\Delta u=510\,\, \AA$. The probability cloud of this state is roughly shaped like a probability ring centered on the origin of the wormhole surface, because the two maxima of the probability density function are very close. Therefore, here the electron is more likely to be found between the wells, i.e. around $u=0$.

The first excited state for $n=10$ (dotted blue line) presents also a probability density function in the form of two Gaussian function, with their maximum located at $u = -328.5\,\, \AA$, and the other located at $u = 328.5\,\, \AA$. The half-height width of each of the Gaussian functions is $\Delta u=1,348.5\,\, \AA$ (see the Fig.\ref{g4} d.2)). Therefore, the probability cloud of the first excited state has the form of two probability rings centered at $u=-328.5\,\, \AA$ and $u=328.5\,\, \AA$ of width $\Delta u=1,348.5\,\, \AA$ each. These probability rings are very close to the origin of the wormhole's surface ($u=0$). The large width of these probability rings show that this state is weakly bound.

The figure \ref{g4} e) shows the effective potential, for $n=20$, in the form of two wells, with its minimuns located at $u=-71\,\, \AA$ and $u=71\,\, \AA$, where its energy are $-228.95$ meV. Here there are two hybrid states, whose energies are $-22.63$ and $-6.72$ meV. The probability density functions of the ground state (red dotted line) has the form of two practically overlapping Gaussian functions with their maximum separated from $u=-70\,\, \AA$ and $u=70\,\,\AA$. Here the width at half height is defined as if there was a single peak due to the proximity between them, so it is not possible to visualize two probability rings, but only one, of the width $\Delta u=206\,\, \AA$, located at $u=0$. So the electron can be found both at the ends of the wormhole and inside it.

The first excited state (blue dotted line) is shown in figure \ref{g4} f), whose probability density takes the form of two Gaussian functions centered on $u=-75\,\, \AA$ and $u=75\,\, \AA$, and the width of each of these Gaussian functions is $\Delta u=83\,\, \AA$. Again, taking into account the angular symmetry of the wormhole, the probability density function of the first excited state (blue dotted line) takes the form of two probability rings located at $u=-75$ and $u=75\,\, \AA$, with each ring having a width of $\Delta u=84\,\, \AA$. Then, the electron is equally likely to be found at the ends of the wormhole.

For $n=40$, in figure \ref{g4} g), the effective potential has two wells located at $u=-71$ and $u=71\,\, \AA$, and its minima have energy equal to $-912.6$ meV. Two bound states of energy are obtained, whose values are $-63.0$ and $-50.4$ meV. The two peaks of the probability density function are at at $u=-71$ and $u=71\,\, \AA$, with the width of $\Delta u=40\,\, \AA$, each one (see the figure \ref{g4} h)). The probability density function peaks for the first excited state have the same location as the ground state function peaks, however its width is $38\,\, \AA$. Therefore, two probability rings, for each bound state, can be visualized symmetrically about the origin of the wormhole. 

It is worth noting that, for $n=40$, the probability density function of the ground state indicates that the electron has the same possibility of being in each of the wells, the same goes for the probability density function of the first excited state, as the two wells are indistinguishable this system has bilateral symmetry, so an electron located in one of the wells can tunnel to the other performing a periodic motion with a frequency given by $f=(E_{1}-E_{0})/h=3 $ THz \cite{novaes}.

Note that the bound states obtained for the GEB wormhole, $n=40$, are more bound than for generalized wormholes for $n<40$, which is reasonable, since the wells become deeper due to the increase in the curvature effect at the edges of the wormhole surface. 

\begin{figure*}
    \centering
    \includegraphics[scale=1.0]{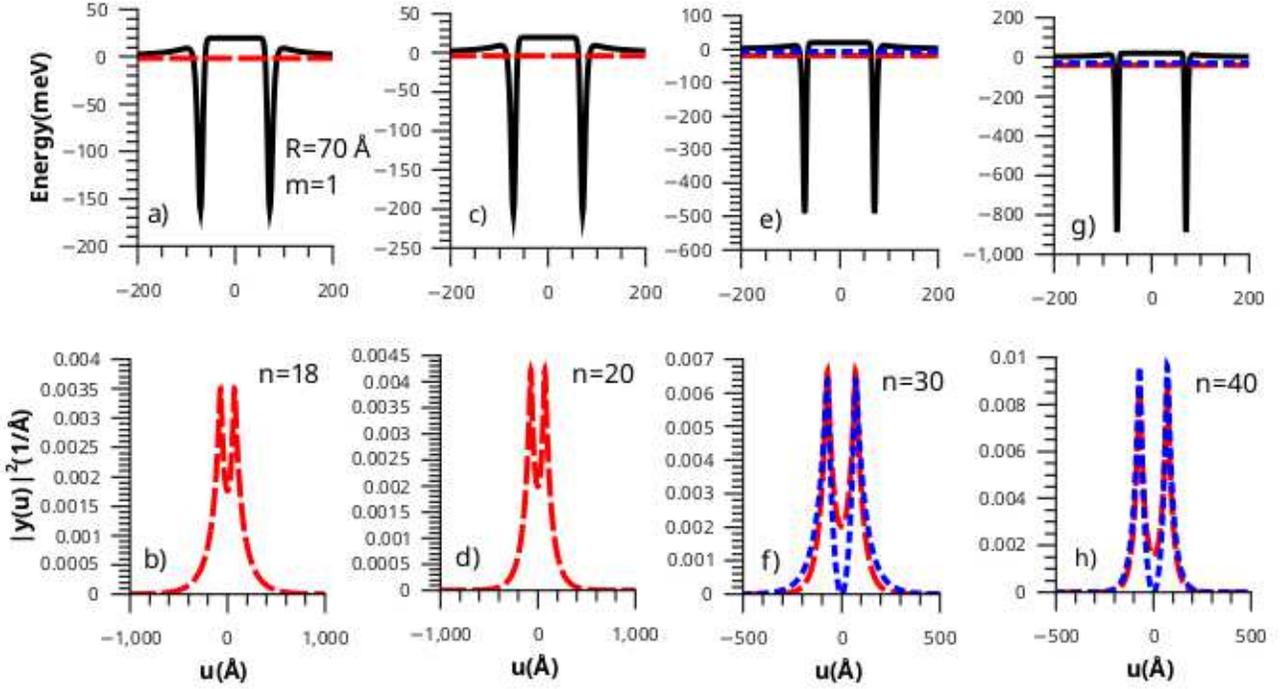}
    \caption{The bound states and their probability densities for a generalized Ellis-Bronnikov wormhole with radius $R=70$ \AA, \ $m=1$ and $m^{*}=0.03m_{0}$. The solid black line represents the effective potential for: a) $n=18$, c) $n=20$, e) $n=30$ and g) $n=40$. The dashed red and dotted blue lines correspond correspond to the ground state and first excited state and their probability densities, respectively.}
    \label{g5}
\end{figure*}

As stated earlier, when the angular momentum is taken into account, $m=1$, the induced potential by the surface wormhole, given by $V_\textrm{ind}(u)$, becomes more relevant than the potential of da Costa $V_\textrm{dC}(u)$, so the effective potential although it takes the form of a potential barrier, for some values of $n$ for GEB wormholes. The effective potential in the form of a double well starts its formation for the GEB wormhole for $n=6$, however, they are very shallow. The first bound state is only obtained for the GEB wormhole for $n=18$, which is shown in figure \ref{g5} a), so the effective potential, represented by solid black line, also presents two wells, with their minima located at $u=-71$ and $u=71 \, \, \AA$, and with energy values equal to $-161.6$ meV. A single state is observed with energy equal to $-1.61$ meV. Its probability density function is shown in figure \ref{g5} b). The probability density function has the form of two almost overlapping Gaussians. Their maximums are located at $u=-72$ and $u=72\,\, \AA$. As the probability density function peaks are very close together, the half-height width of each peak loses resolution and is seen as a single peak of width $\Delta u=206\,\, \AA$. Then, only one probability ring is observed at the center of the wormhole, which is the region where the electron is most likely to be located. 
 
 For $n=20$, also a single state is observed with energy equal to $-3.85$ meV. The effective potential minima are located at $u=-71$ and $u=71\,\, \AA$, with energy value equal to $-204.84$ meV. The figures \ref{g5} c) and d) show the effective potential, the level energy and the probability density function. Here, for $n=20$, the probability density function is similar to the one shown for $n=18$ (see figure \ref{g5} d)), however the peaks of the probability density function are narrower, with $\Delta u=100 \, \, \AA$, Although the peaks are narrower, the resolution of the two probability rings is not very clear, therefore, a single probability ring is observed at the center of the wormhole surface.
 
 The figure \ref{g5} e), for $n=30$, shows two bound states with equal energies $-19.5$ and $-7.53$ meV. The effective potential have two minima with value of $-491$ meV (solid black line). The probability density functions of the two confined states are practically similar. That is, two probability rings located at $u=-71$ and $u=71$ \AA, with a width of $\Delta u=56$ \AA, as shown in figure \ref{g5} f). 
 As the probability density functions have similarities to those discussed for the wormhole for $m=0$ and $n=40$, here for $m=1$ and $n=30$, here too the electron can perform oscillatory motion with
 frequency $f=2.9$ THz. 
 
 The effective potential, for the surface of an GEB if $n=40$, has two bound states with energies equal to $-41.6$ and $-30.6$ meV.  The minima of the potential well are $-887.7$ meV (see figure \ref{g5} g)). The probability density function of the two states exhibits two peaks. The width of the ground state peaks is $\Delta u=40\,\, \AA$, while the first excited state is $\Delta u=38\,\, \AA$. Hence, the ground state probability cloud consists of two rings of width $40\,\, \AA$ one at $u=-71\,\, \AA$ and the other at $u=71\,\, \AA$. Whereas the first excited state consists of two rings of width $38\,\, \AA$ located at the same positions as the ground state, as shown in figure \ref{g5} h). Here too the electron can oscillate from one well to another with frequency $f=2.7$ THz.  
 
 When changing the radius $R$ of the wormhole there is no qualitative change, however, quantitatively there are changes in the values of energy levels. As $R$ decreases the effective potential becomes deeper, due to the increasing effect of the curvature of the wormhole surfaces, so the energy levels become more confined. And the increasing in the value of $R$ has the opposite effect, the effective potential becomes less deep, and the levels become less confined.
 
The table below shows the oscillation frequencies for electrons oscillating between the ends of the wormhole ($n=40$), for some values of $R$. Note that as $R$ increases, the frequency of oscillation decreases.
 \begin{center}
 \begin{tabular}{|c|c|c|c|c|c|}
		\hline 
	R	& $n$ & $m$ & $f=\frac{\Delta E}{h}$ & $m$ &  $f=\frac{\Delta E}{h}$\\ 
		\hline 
	50 \AA	& 40 & 0 & 6.0 THz & 1 & 5.2 THz \\ 
		\hline 
	70 \AA	& 40 & 0 & 3.0 THz &  1& 2.7 THz \\ 
		\hline 
	100 \AA	& 40 & 0 & 1.5 THz & 1& 1.2 THz \\ 
		\hline 
	200	\AA& 40 & 0 & 0.4 THz & 1 & 0.3 THz  \\ 
		\hline 
\end{tabular}
 \end{center}


\section{Final Remarks}
In this work we study the electron interacting with the surface of a Generalized Ellis-Bronnikov wormhole (GEB) made of graphene, via effective mass approximation. For that, an effective potential that confines the electron on the surface of wormhole GEB is obtained. This effective potential is the combination of an induced potential by the surface, $V_\textrm{ind}(u)$, and the da Costa potential, $V_\textrm{dC}(u)$, which emerges from the squeezing of the electron to the surface of the wormhole. The induced potential by the surface wormhole GEB is essentially repulsive, whereas the da Costa potential is attractive. 
 
In the absence of orbital angular momentum ($m=0$), the da Costa potential predominates over the induced potential, which causes the effective potential to take the form of a potential well for $n=2$, or a double potential well for $n>2$. Solving Eq.(\ref{39}) for these potentials considering $R=70$ \AA \, one bound state is obtained for $n=2$, and two states are obtained for $n>8$, the latter are hybrids.

The effective potential is drastically altered when the orbital angular momentum is taken into account ($m\neq 0$), since the induced potential becomes more relevant than the da Costa potential. Although the induced potential, repulsive, has relevance in relation to the da Costa potential, attractive, a bound hybrid state for $m=1$, $R=70$ \AA \ and $n>16$ is obtained, and two hybrid states for $n>24$. 
 
A frequency of an electron oscillating between these states was estimated of the order of THz. This frequency can be used as a way of characterizing the system itself.
 
The effective potential does not show qualitative change when the radius of the system is modified, however, the values of the energy levels can be changed. For example, for smaller $R$ the effective potential depth increases, and although the number of states is not changed, these states assume low energy values, so the radiation emitted by the electron when transiting from one well to the other can also be modified. Here it can be inferred that the electronic oscillation frequency can be related to the size of the system.

A natural extension of this work is to analyze these results when we have the action of external fields (constant electric and magnetic fields). Another interesting possibilities are the study of the thermodynamic properties of the present system as well as the investigation of a small twist between the upper and lower graphene sheets.

\section{Acknowledgments*}

RNCF would like to thank CNPQ grant 312384/2018-1 and TFS would like to thank CAPES for financial support.

\appendix
	
	\section{Mean and Gaussian curvatures}
	
The GEB wormhole has symmetry around the $z$-axis, the parametrization can be given by \cite{manfredo, spivak}
\begin{align}
    x(u,\phi)&=f(u)\cos \phi, \label{8} \\
    y(u,\phi)&=f(u)\sin \phi, \label{9} \\
    z(u)&=h(u) \label{10},
\end{align}
where $f(u)=\left(u^n+R^n\right)^{1/n}$ and $h(u)$ is a homeomorphic continuous function such that $dh$ is one-to-one. For the moment, we don't worry about knowing the function $h(u)$ that makes the GEB wormhole surface regular, but its derivatives must be known so that we can calculate the curvatures. The vectors that constitute any basis for the space tangent to the surface of the GEB wormhole are \cite{spivak}
\begin{align}
    \vec{r}_u(u, \phi)&= u^{n-1}\left(u^n+R^n\right)^{1/n-1}\,\hat{\rho}+h'\, \hat{k}, \label{11} \\
    \vec{r}_\phi(u, \phi)&=\left(u^n+R^n\right)^{1/n}\hat{\phi}, \label{12}
\end{align}
such that the necessary condition for the GEB wormhole surface to be regular leads to
\begin{equation}
    \left(h'(u)\right)^2=1-u^{2n-2}\left(u^n+R^n\right)^{2/n-2}. \label{13}
\end{equation}
With (\ref{11}) and (\ref{12}), the unit normal vector to GEB wormhole surface is
\begin{equation}
    \hat{n}(u, \phi)=u^{n-1}\left(u^n+R^n\right)^{1/n-1}\, \hat{k}-h'\, \hat{\rho}. \label{14}
\end{equation}
We have also
\begin{align}
    \vec{r}_{uu}(u, \phi)&=(n-1)R^nu^{n-2}\left(u^n+R^n\right)^{1/n-2}\, \hat{\rho}+h''\, \hat{k}, \label{15} \\
    \vec{r}_{u\phi}(u, \phi)&=u^{n-1}\left(u^n+R^n\right)^{1/n-1}\hat{\phi}, \label{16} \\
    \vec{r}_{\phi \phi}(u, \phi)&=-\left(u^n+R^n\right)^{1/n}\hat{\rho}. \label{17}
\end{align}
where we have $\vec{r}_i=\partial \vec{r}/\partial x^i$ and $\vec{r}_{ij}=\partial^2\vec{r}/\partial x^i\partial x^j$.

From (\ref{15}), (\ref{16}) and (\ref{17}), the coefficients of the second form of (\ref{8}), (\ref{9}) and (\ref{10}) will be
\begin{align}
    h_{uu}(u)&=\vec{r}_{uu}\cdot \hat{n}=h''u^{n-1}\left(u^n+R^n\right)^{1/n-1} \nonumber \\
    &\hspace{8mm}-h'(n-1)R^nu^{n-2}\left(u^n+R^n\right)^{1/n-2}, \label{18} \\
    h_{u\phi}(u)&=\vec{r}_{u \phi}\cdot \hat{n}=0, \label{19} \\
    h_{\phi\phi}(u)&=\vec{r}_{\phi\phi}\cdot \hat{n}=h'\left(u^n+R^n\right)^{1/n}. \label{20}
\end{align}
When $n=2$, we get $h_{uu}<0$ and $K<0$ (gaussian curvature), and that $h_{u\phi}$ and $g_{u\phi}$ are all zero for any $n$. For this to continue to be respected for all $n=2,4,6, ...$ , it is necessary to impose $h'<0$. Therefore (\ref{18}), (\ref{19}) and (\ref{20}) become
\begin{align}
    h_{uu}(u)&=-(n-1)R^nu^{n-2}\left(u^n+R^n\right)^{1/n-2} \nonumber  \\
    &\hspace{8mm} \times \left(1-u^{2n-2}\left(u^n+R^n\right)^{2/n-2}\right)^{-1/2}, \label{21}  \\
    h_{u\phi}(u)&=0, \label{22} \\
    h_{\phi\phi}(u)&=\left(u^n+R^n\right)^{1/n}\left(1-u^{2n-2}\left(u^n+R^n\right)^{2/n-2}\right)^{1/2}. \label{23}
\end{align}
The mean and gaussian curvatures are defined by
\begin{align}
    H&=\frac{1}{2}\left(\frac{h_{uu}g_{\phi\phi}-2h_{u\phi}g_{u\phi}+h_{\phi\phi}g_{uu}}{g_{uu}g_{\phi\phi}-g_{u\phi}^2}\right), \label{27} \\
    K&=\frac{h_{uu}h_{\phi\phi}-h_{u\phi}^2}{g_{uu}g_{\phi\phi}-g_{u\phi}^2}, \label{28}
\end{align}
and now can be straightforwardly calculated yielding
\begin{align}
    H(u)&=\frac{1}{2}\left(1-u^{2n-2}\left(u^n+R^n\right)^{2/n-2}\right)^{-1/2} \biggl(\left(u^n+R^n\right)^{-1/n}- \nonumber \label{29} \\
    &\hspace{6mm}(n-1)R^nu^{n-2}\left(u^n+R^n\right)^{1/n-2} \\ \nonumber 
    &\hspace{8mm}-u^{2n-2}(u^n+R^n)^{1/n-2}\biggr)\,, \\
    K(u)&=-(n-1)R^nu^{n-2}\left(u^n+R^n\right)^{-2}. \label{30}
\end{align}	


\end{document}